\documentclass{article}    

\usepackage{graphicx}

\title{\textbf{NuRule (4) and the 3-Level Atom}}  
\author{Richard Mould\footnote{Department of Physics and Astronomy, State University of New York, Stony Brook,
\mbox{New York} 11794-3800, http://ms.cc.sunysb.edu/\~{}rmould} }  
\date{}    
   
\begin{document}             

\maketitle              

\begin{abstract}

When a weak decay competes with a strong decay in a 3-level atom, some mechanism is necessary to occasionally stop the
strong decay so the weak decay can be completed.   Rule (4) provides that mechanism.  Using this rule, a weak photon is
emitted at the correct time for both the V and $\Lambda$ configurations, as well as for the two cascade
configurations.     
   
\end{abstract}

\section*{Introduction}
	A single three-level atom has a ground state 0 and excited states 1 and 2.  The ``strong" decay (1-0) has a lifetime of
about $10^{-8}s$, and the ``weak" decay (2-0) has a lifetime of about 2s.  The atom is exposed to two laser beams, where
the first is tuned to the transition 0-1 and the second is tuned to 0-2.  

		The atom will oscillate rapidly between the ground state and the first excited state for a period of time, giving off a
visible light.  It will then stop radiating and become dark for a period of time, after which it will resume the visible
radiation.  This light-dark \emph{florescent pulsing} is thought to be due to the atom being trapped in the weak state 2
during the dark time \cite{HD, AS}.  The florescent radiation is said by Dehmelt to be suspended for a time because the
atom is temporarily ``shelved" in state 2.    

The question is: What is the quantum mechanical mechanism that causes this shelving?  It is shown below that that 
mechanism is nuRule (4).  The first of the nuRules provides for the existence of a stochastic trigger, and the other three
are textually included in the following.  All four nuRules are listed in a previous paper \cite{RM3}.

\section*{The Strong Interaction by Itself}
Imagine that only the first laser (0-1)  is turned on at time $t_0$.  The first two cycles of the strong interaction will
then take the form
\begin{eqnarray}
\Phi(t =t_0) &=& A_0\nonumber\\
\Phi(t \ge t_0)&=& A_0 + A_1 + A_0\phi + A_1\phi + A_0\phi\phi + \hspace{.1cm}. . .
\end{eqnarray}
where $A_0$ is the ground state of the atom and $A_1$ is its excited state.  In the second component, the atom has absorbed
a photon from the laser beam, and in the third component it has returned to the ground state and emitted a photon
$\phi$.  The fourth and fifth components complete another cycle.  The process repeats indefinitely.  Laser photons are not
shown.  

After $t_0$ the second component acquires amplitude as a result of current flow from the first component, then the third
component, then the fourth, etc.  One can imagine a pulse of states that proceeds from left to right in eq.\ 1, preserving
normalization.  This results in a steady stream of photons leaving the atom every $10^{-8}$ s, giving visible
radiation.  

Add a detector $\Phi = A\otimes D$ and the first two cycles will become 
\begin{equation}
\Phi(t \ge t_0) = (A_0 + A_1)\otimes D_0 + 
\underline{A}_0\underline{D}_1 (+) \underline{A}_1\underline{D}_1 (+)\underline{A}_0\underline{D}_2 (+)
\underline{A}_1\underline{D}_2 (+)
\hspace{.1cm} ...
\end{equation}
where the detector's ground state is $D_0$.  After the first component \mbox{$(A_0 + A_1)\otimes D_0$} the atom and the
detector become entangled through the radiation field.  Emitted photons are assumed to be immediately captured by the
detector, so one photon has been recorded by $D_1$ and two photons by $D_2$. 

NuRule (2) requires that every component after the first component in \mbox{eq.\ 2} will contain \emph{ready} states.  The
underline  in \mbox{eq.\ 2} identifies these states.  They are the basis states of a collapse, and they gain that status
when the component they occupy is incoherent with the other components - generally through environmental decoherence.  

\vspace{0.3cm}
\textbf{nuRule (2)}: \emph{If the Hamiltonian gives rise to new components that are locally incoherent with, and are
discontinuous with the old components or with each other, then all states that are included in the new components will be
ready states.}
\vspace{0.3cm}

In the case of eq.\ 2, all the new components are environmentally decoherent because of the macroscopic nature of the
detector. The parentheses around the + sign means that probability current cannot flow between the indicated components
because of nuRule (4), which says that current cannot pass between components that both contain ready states of the same
object.  

\vspace{0.3cm}
\textbf{nuRule (4)}: \emph{A transition between two components is forbidden if each is an entanglement containing a ready
state of the same object.}
\vspace{0.3cm}

Because of rule (4), the process in eq.\ 2 is stalled until there is a stochastic hit on $\underline{A}_0\underline{D}_1$
at time
$t_{sc1}$.  That hit will make $A_0$ and $D_1$ \emph{realized} states and will reduce all other components to zero
according to nuRule (3).  A state that is not ready is called realized.  

\vspace{0.3cm}
\textbf{nuRule (3)}: \emph{If a component containing ready states is stochastically chosen, then all of the states in that
component will become realized, and all other components will be immediately reduced to zero.}
\vspace{0.3cm}

	The above collapse also disentangles the atom from the detector because it projects into exact eigenstates of the
measurement,  so at the time $t_{sc1}$ of the stochastic choice we have
\begin{equation}
\Phi(t \ge t_{sc1} > t_0) = (A_0 + A_1)\otimes D_1 + \underline{A}_0\underline{D}_2 (+) \underline{A}_1\underline{D}_2 (+)
\hspace{0.1cm}. . .
\end{equation}
where the new components (underlined) are all zero at $t_{sc1}$.  The second component thereafter gains amplitude due to
current flow from the first component.  However, no current will flow into the third component in eq.\ 3 because of
\mbox{rule (4).}  After a second stochastic hit at time $t_{sc2}$ the state of the system becomes
\begin{equation}
\Phi(t \ge t_{sc2} > t_{sc1} > t_0) = (A_0 + A_1)\otimes D_2 + \underline{A}_0\underline{D}_3 (+)
\underline{A}_1\underline{D}_3 (+)
\hspace{0.1cm}. . .
\end{equation}
and so forth.  It is clear that each new cycle must await a stochastic hit before it can go to the next cycle.  Each hit 
is followed by a renewal and a continuation of the process.

\section*{A Competitor}
Since all the current from the initial state in eq.\ 2 flows into the second component, it is certain that there will be a
stochastic hit on the second component at some time $t_{sc1}$.  However, if there is a competing interaction in parallel
with \mbox{eq.\ 2}, then the current going into the second component will fall short of the amount necessary to
\emph{guarantee} a hit on that component.  In that case, the interaction in eq.\ 2 will be stalled until the competing
interaction has run its course - however long that may take.  This conclusion does not depend on the competitor being
strong or weak relative to the interaction in eq.\ 2.  The basic mechanism of the florescent pulse is thereby in place.  

Imagine that only the second laser (0-2)  is turned on at time $t_0$.  The first three cycles of the weak interaction will
then take the form
\begin{eqnarray}
\Phi(t =t_0) &=& A_0\nonumber\\
\Phi(t \ge t_0)&=& A_0 + A_2 + A_0\phi' + A_2\phi' + A_0\phi'\phi' + A_2\phi'\phi' \hspace{.1cm}. . .
\end{eqnarray}
Adding the detector gives
\begin{equation}
\Phi(t \ge t_0) = \{A_0 + A_2 + A_0\phi' + A_2\phi' + A_0\phi'\phi' + A_2\phi'\phi' \hspace{.1cm}. . .\} \otimes D_0
\end{equation}
because the detector does not interact with the primed photons that are products of the 2-0 decay.

\section*{Both at Once}
If the first and second lasers are turned on together, then eqs. 2 and 6 will be competing processes.  Starting with
$\Phi(t = t_0) = A_0\otimes D_0$, we get 
\begin{eqnarray}
\Phi(t \ge t_0) &=& A_0\otimes D_0\\
&+& A_1\otimes D_0 + \underline{A}_0\underline{D}_1 (+)\hspace{0.05cm} ... \hspace{0.2cm}\mbox{from eq.\ 2}\nonumber\\
&+& \nonumber\{A_2 + A_0\phi'\}\otimes D_0 + ... \hspace{0.2cm}\mbox{from eq.\ 6}
\end{eqnarray}
where both the second and the third rows must share the current flowing from the first row, and where the current received
by the second row will be much greater than the current received by the third row.  Therefore, in most cases
$\underline{A}_0\underline{D}_1$ in the second row will be stochastically chosen at time $t_{sc1}$ giving 
\mbox{$\Phi(t = t_{sc1} > t_0) = A_0\otimes D_1$} which begins the process all over again.
\begin{eqnarray}
\Phi(t \ge t_{sc1} > t_0) &=& A_0\otimes D_1\\
&+& A_1\otimes D_1 + \underline{A}_0\underline{D}_2 (+)\hspace{0.05cm} ... \nonumber\\
&+& \{A_2 + A_0\phi'\}\otimes D_1 + ... \nonumber
\end{eqnarray}
most probably resulting in another stochastic hit in the second row giving 
\begin{equation}
\Phi(t = t_{sc2} > t_{sc1} > t_0) = A_0 \otimes D_2
\end{equation}
It is the rapid repetition of eqs.\ 7-9, that produces the radiation observed during the florescent ``on" time of the
3-level atom.  

	However, the third row in eq.\ 8 competes with the second row for current.  This means that the probability of a
stochastic hit on $\underline{A}_0\underline{D}_2$ in the second row of eq. 8 might not occur; in which case, the square
modulus of the component will stop increasing.  When that happens $\underline{A}_0\underline{D}_2$ will become a phantom
component, lying dormant for a time.  Its only function during this dormancy is to stall the strong decay.  The third row
will then continue the evolution by itself.  It will take a (comparatively) long time to execute the cycle from $A_0$ in
the first row of eq.\ 8 back to $A_0$ in the third row, resulting in the ``dark time" of the 3-level florescent pulse.  

When that execution is complete, current can again flow into $A_1\otimes D_1$ in the second row, and then to the dormant
phantom component $\underline{A}_0\underline{D}_2$.  This will reactivate the phantom, driving its square modulus closer to
1.0 than was previously achieved.  A stochastic hit accompanying this new surge of current will generally  start a new
period of florescence; although it is possible that there will be no stochastic hit, and instead, a second dark period
will begin.  Following such a second dark period, the phantom component will be given still another chance to be
selected.  The probability is 1.0 that $\underline{A}_0\underline{D}_2$ will eventually be stochastically chosen,
initiating a period of florescence that begins the full cycle over again.

\section*{The V and $\Lambda$ Configurations}
	The above arrangement of energy levels is called the V configuration because levels 1 and 2 are above the ground level
0.  It is a characteristic of this case that the dark period \emph{ends} with the emission of a weak photon \cite{PLK}. 
This is confirmed in eq.\ 8 because $\phi'$ appears in the last term in the third row, after the dark period is complete.

In the $\Lambda$ configuration levels 1 and 2 are below the ground level 0.  In this case eq.\ 1 is 
\begin{eqnarray}
\Phi(t =t_0) &=& A_0\nonumber\\
\Phi(t \ge t_0)&=& A_0 + A_1\phi + A_0\phi + A_1\phi\phi + A_0\phi\phi + \hspace{.1cm}. . .
\end{eqnarray}
and eq. 2 becomes
\begin{equation}
\Phi(t \ge t_0) = A_0\otimes D_0 + 
\underline{A}_1\underline{D}_1 (+) \underline{A}_0\underline{D}_1 (+)\underline{A}_1\underline{D}_2 (+)
\underline{A}_0\underline{D}_2 (+)
\hspace{.1cm} ...
\end{equation}
The equivalent of eq.\ 5 is then
\begin{equation}
\Phi(t \ge t_0)  = A_0 + A_2\phi' + A_0\phi' + A_2\phi'\phi' + A_0\phi'\phi' +  \hspace{.1cm}. . .
\end{equation}
making eq.\ 6 
\begin{equation}
\Phi(t \ge t_0) = \{A_0 + A_2\phi' + A_0\phi' + A_2\phi'\phi' + A_0\phi'\phi' +  \hspace{.1cm}. . .\} \otimes
D_0
\end{equation}
Combining these two processes, we get
\begin{eqnarray}
\Phi(t \ge t_0) &=& A_0\otimes D_0\\
&+& \underline{A}_1\underline{D}_1 (+) \hspace{0.05cm} ... \hspace{0.2cm}\mbox{from eq.\ 11}\nonumber\\
&+& \nonumber\{A_2\phi' + A_0\phi'\}\otimes D_0 + ... \hspace{0.2cm}\mbox{from eq.\ 13}
\end{eqnarray}
where both the second and third rows must share the current flowing from the first row. 

In this case the photon $\phi'$ appears in the first term in the third row of \mbox{eq.\ 14}, so  the dark period of
the
$\Lambda$ configuration \emph{begins} rather than ends with the emission of a weak photon.

\section*{The Cascade Configurations}
Consider first that state 2 (the weak photon) is at a higher energy than the ground state 0, and state 1 is at a lower
energy than ground.  If both lasers are turned on together, the sum will combine eqs.\ 6 and 11.
\begin{eqnarray}
\Phi(t \ge t_0) &=& A_0\otimes D_0\\
&+& \underline{A}_1\underline{D}_1 (+) \hspace{0.05cm} .. \hspace{0.2cm}\mbox{from eq.\ 11}\nonumber\\
&+& \nonumber\{A_2 + A_0\phi'\}\otimes D_0 + .. \hspace{0.2cm}\mbox{from eq.\ 6}
\end{eqnarray}
Giving another case in which the weak photon is emitted at the \emph{end} of the dark period. 

Finally, let state 2 (the weak photon) be at a lower energy than the ground state 0, and let state 1 be at a higher energy
than ground.  If both lasers are turned on together, the sum will combine eqs.\ 2 and 13.
\begin{eqnarray}
\Phi(t \ge t_0) &=& A_0\otimes D_0\\
&+& A_1\otimes D_0 + \underline{A}_0\underline{D}_1 (+) \hspace{0.05cm} .. \hspace{0.2cm}\mbox{from eq.\ 2}\nonumber\\
&+& \nonumber\{A_2\phi' + A_0\phi'\}\otimes D_0 + .. \hspace{0.2cm}\mbox{from eq.\ 13}
\end{eqnarray}
Giving another case in which the weak photon is emitted at the \emph{beginning} of the dark period.

\section*{Original Rules}
	Prior to formulating these nuRules, the author proposed other rules (1-4) that confined the basis reduction states to
observer brain states \cite{RM5}.  According to these rules, a quantum mechanical state will not collapse unless an
observer interacts with it; so ``objective" measurements are not possible.  When these original rules are applied to the
3-level atom, the behavior of the atomic/radiation system is the same as above.  Florescent pulsing is therefore fully
accounted for by the original rules; but  in this case, pulsing does not occur unless an observer is present.  Without an
observer, the system will reach a steady ``unpulsed" equilibrium as a result of the uninterrupted Schr\"{o}dinger process
acting alone.  

This may seem like an unacceptable result.  But the fact is that there is no empirical way to determine whether or not
there are florescent pulses without an observer on the scene.  One might suggest that the system be recorded on video tape
and observed much later, hoping thereby to disengage the observer from the detector.  But that won't work.  Because of
non-local correlations, the result would be the same as it is when an observer looks directly at the detector.  The result
would confirm the existence of florescent pulsing.  It would not tell us what happens when there is no observation at
all.  Therefore, we cannot empirically affirm or deny the existence of pulsing when there is no observer. 

The idea that florescent pulsing depends on the presence of an observer might be counterintuitive, leading one to abandon
the original rules in favor of the nuRules that give a more acceptable result.  But since there can be no empirical
evidence one way or the other, and since there is no currently established theory of measurement that would settle the
matter, it is wise to keep these options open.  Maybe florescent pulses are a consequence of observer/system interaction. 
We are not accustomed to thinking that observers can have this kind of influence on physical systems under observation. 
But customary thinking for the last 400 years has placed the consciousness of an observer \emph{outside} of the universe,
and that's not realistic either.  Maybe, when we finally integrate conscious observers into the quantum
mechanical universe, we will discover that we (conscious creatures) play a role in Nature that we might not have otherwise
suspected.  Therefore, we should wait and see which of these rule sets (the old rules or the nuRules) is most convincingly
incorporated into a wider theoretical understanding.  

It is important to realize that even the nuRules produce counterintuitive results; namely, that florescent pulses require
the existence of an atom/detector interaction.  Without a macroscopic detector present to make the measurement, the atomic
system governed by the nuRules will not pulsate.  The atom and laser field by itself cannot produce ready states because
the participating atomic levels are not mutually incoherent; so nuRule (4) will not be able to eliminate second order
transitions among these levels.  Because of this, the strong interaction cannot be brought to a halt while the weak
interaction completes its cycle.  This result would probably please Niels Bohr, for it ties the existence of an observable
atomic phenomena directly to the macroscopic detector that is used to detect it.

\end{document}